# LLC type galvanic isolated resonant converter

Antonio Brajdić
Faculty of Electrical Engineering and Computing
University of Zagreb, HR-10000 Zagreb, Croatia
E-mail: Antonio.brajdic@fer.hr

*Abstract*— **Resonant converters are often being used for high power and high voltage applications to achieve high efficiency, high power density and low EMI. In this paper, we will use a resonant converter for a completely different application. We will take advantage of the Zero Voltage Switching (ZVS) and operating in the vicinity of the series resonance to design a low power and ultra-low noise galvanic isolated half bridge resonant DC-DC converter for powering a medical amplifier. Under lower power we consider up to 6W (500mA @ 12V). The DC-DC converter must meet the safety requirements of the standard IEC 60601.**

## I. INTRODUCTION

There are many resonant-converter topologies, and all operate in the same way. A resonant converter has 3 main parts: a switching circuit, a resonant tank circuit and a rectification. Among all different topologies of resonant converters, there are two basic types: a series resonant converter and a parallel resonant converter. A more improved type is a series-parallel resonant converter, SPRC. This type comes in two version, LLC and LCC (Figure 1). The first one uses two capacitors and an inductor, and the second one uses two inductors and a capacitor. The LLC version is more common because two inductors can often be integrated into a one physical inductor. LLC converters can regulate the output over wide line and load intervals with a quite small variation of the switching frequency, while maintaining an excellent efficiency. It can also achieve the zero-voltage switching (ZVS) over the entire operating range.

The switching circuit is usually made of MOSFETS, which generates a square wave on the output. Both are driven with 50% duty time with some dead time between cycles to achieve the zero voltage switching, ZVS.

In the LLC configuration, a resonant tank circuit is made of a capacitor and two inductances – a series resonant inductance and a transformer magnetizing inductance. A resonant tank has multiple roles in this type of converter. With changing the switching frequency, impedance of the resonant tank changes and this changes the amount of energy going to the load. So, it is possible to say that the resonant tank and a load act like a voltage divider. The resonant tank impedance is minimum at the resonant frequency of the resonant network. This means that the resonant circuit will attenuate all signal frequencies except one on the resonant frequency. As a square wave enters the resonant circuit and the fundamental harmonic of the square wave signal is a sine wave, after attenuation of all

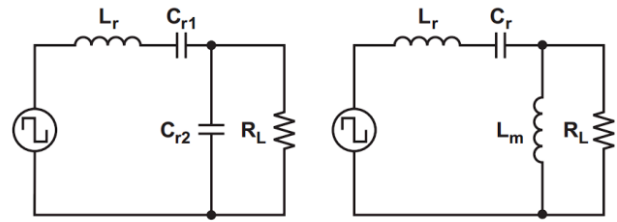

*Figure 1 (a) LCC configuration,    (b) LLC configuration [7]*

higher order harmonics, we will get a pure sinusoidal current inside the resonant tank. This is also called FHA, a first harmonic approximation. While operating in the inductive region, the tank current lags the voltage and makes ZVS possible to achieve.

Recertification is a part of the secondary side of the converter. It is made from two diodes in a center tapped configuration of the transformer. Output capacitors are added to smooth the rectified voltage and current. If needed, an additional LC filter can be added to the output to make the smoother DC output.

LLC filter has two resonant frequencies, depending on the load. If no load is present, the frequency is $f_p = \frac{1}{2*pi*\sqrt{(L_r+L_m)*C_r}}$ (1). At a load short circuit, the resonant frequency is a series resonant frequency $f_0 = \frac{1}{2*pi*\sqrt{L_r*C_r}}$ (2) because $L_m$ is shorted by the load. Hence, during the time the switching frequency varies between $f_p$ and $f_0$ depending on the load.

Operation of the LLC converter can be characterized by the relationship of the switching frequency to the series resonant frequency. Converter can operate at, below and above the resonance.

When operating at the resonance, the resonant tank current is in phase with the load current. It is possible to achieve ZVS, but there is only a single operating point of the converter which is not optimal for the regulation.

When operating below the resonance, the impedance of the resonant tank is capacitive, and the current leads the voltage. That means that the resonant tank current falls to the magnetizing current before the MOSFET turns off and makes ZVS difficult to achieve.

On the other hand, when operating above the resonance, the resonant tank impedance is inductive. In that case,

resonant current is behind the voltage which makes ZVS easy to achieve.

In conclusion, to achieve ZVS the resonant tank impedance must be inductive and there must be enough energy in the resonant tank. Advantages of ZVS are soft switching on MOSFETs which results with lower power losses and, for our case the more important thing is that ZVS eliminates high current spikes and lowers the EMI noise.

## II. METHODOLOGY

Input and output voltage of the converter are connected by a voltage gain function, also called a transfer function. As said previously, an LLC converter is based on the FHA, first harmonic approximation. As the fundamental harmonic of a square wave is a sine, that means that the current inside of the resonant tank will be sinusoidal. But this is only valid when the converter is operated in the vicinity of the resonant frequency. Going away from the resonant frequency, more higher frequency harmonics are included, and the FHA is no longer applicable. This is the reason why the converter should work in the vicinity of the resonant frequency. And the voltage gain function is dependent on the working frequency so the voltage gain should be adjusted to maintain the working frequency close to the resonant frequency. Connection between the input and output voltage is given by this function $V_o = M_g * \frac{1}{n} * \frac{V_{in}}{2}$ (3) where the $M_g = \frac{L_n * f_n^2}{[(L_n+1)*f_n^2-1]+j[(f_n^2-1)*f_n*Q_e*L_n]}$ (4) stays for the voltage gain function. $L_n = \frac{L_m}{L_r}$ (5) are two inductances combined into one. $f_n = \frac{f_{sw}}{f_0}$ (6) is the normalized frequency and $Q_e = \frac{\sqrt{L_r/C_r}}{R_e}$ (7) is the quality factor of the series resonant circuit. In the gain function, $L_n$ and $Q_e$ are fixed variables because they are values of physical components and $f_n$ is a control variable. After design is complete, $M_g$ is adjusted by $f_n$. Easiest way to explain that is to look at the plot in Figure 2.

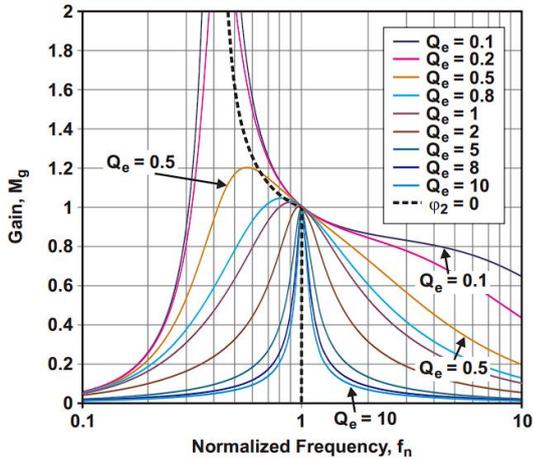

*Figure 2 Voltage gain function for fixed value of $L_n$ [7]*

For given $L_n$ and $Q_e$, $M_g$ represents convex curves in the vicinity of the resonance. In Figure 2 for given $L_n = 5$, we can see $M_g$ for different values of $Q_e$ as $Q_e$ changes with the load. When $R_l$ is opened, $Q_e$ is 0 and $f_{c0} = f_p$. For $R_l$ shorted, $Q_e = \infty$.

As the output voltage is regulated by $M_g$ through controlling $f_n$, $M_g$ is varying in a limited interval for achieving load and line regulation. Line regulation is defined as the maximum output-voltage variation caused by an input voltage variation over a specified range, for a given output load current. Load regulation is defined as the maximum output-voltage variation caused by a change in load over a stated range, usually from no load to the maximum.

To achieve line and load regulation, the converter must work in some range of frequencies which is regulated by values of $M_g$: $M_{g\_min} = \frac{n*V_{o\_min}}{V_{in\_max}/2}$ (8), $M_{g\_max} = \frac{n*V_{o\_max}}{V_{in\_min}/2}$ (9) and $M_{g\_\infty} = \left|\frac{L_n}{L_n+1}\right|$ (10).

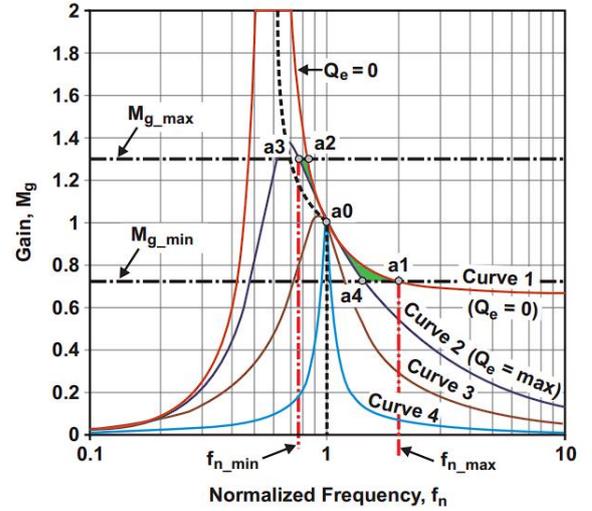

*Figure 3 Recommended design area [7]*

As we can see in Figure 3, $M_{g\_min}$ and $M_{g\_max}$ are horizontal lines and curves changes shape depending on the load. For the *Curve 0* for a no load to the *Curve 2* for a maximum load and higher curves for *overload* condition. We can see that for *Curve 3*, $M_{g\_max}$ is not able to cross the *Curve 3* and converter isn't able to achieve regulation.

One important situation to cover is a short circuit, as in a short circuit condition a big amount of a current is circulating in the circuit. The *Curve 4* in Figure 3 shows a load short circuit gain. It is easy to see that with increasing the switching frequency, the gain drops and the current drops. Another thing is how fast the controller will detect a short circuit condition and how much time will be needed to drop the voltage gain.

For this application, we have chosen the switching frequency of 100kHz. There are multiple reasons for choosing this frequency. One is that EMI testing starts at 150kHz, and another, more important one, is that there are more components developed for this frequency, primarily cores for inductors and transformers, and they are less expensive and switching losses are lower.

Transformer turns ratio is selected from the main voltage gain function (2). As in our case the input voltage is 48V and the output voltage is 12V, *n* should be 2. But for our application we have chosen *n* to be 1.83. The main reason is that with turn ratio of 2, the middle of the voltage gain is set to 90kHz as can be seen from Figure 4. That means that the controller is not working at the resonant frequency. With adjusting n to be 1.83, the middle of the voltage gain is set to 100kHz with +/- 10kHz frequency change with the gain change.

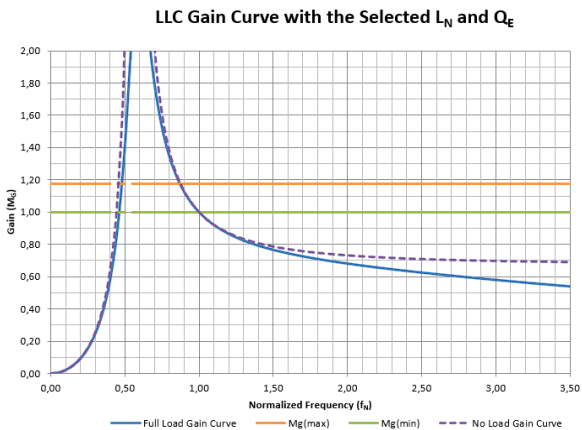

*Figure 4 n = 2*

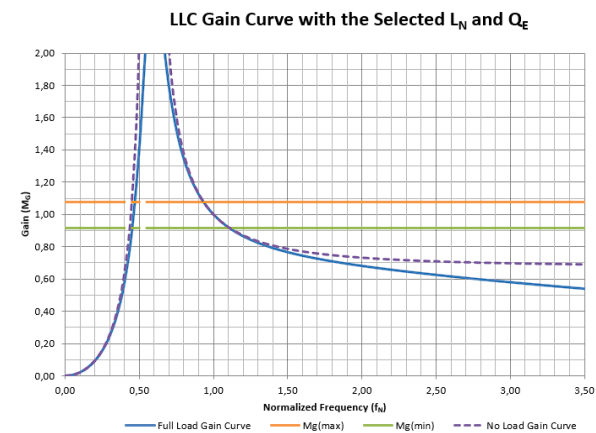

*Figure 5 n = 1.83*

To achieve line and load regulation we need to select $L_n$ and $Q_e$ values because they change the gain frequency curves and must select the proper value for the curve to cross two horizontal lines. The most critical point is the point a3 (Figure 3) which represents the maximum attainable gain which occurs at the maximum load current. This point must be carefully set to work in the inductive region. After $M_{g\_max}$ is chosen we can select $L_n$ and $Q_e$.

In our design we have selected value of $L_n = 2.05$ and $Q_e = 0.36$. The reason for these values of $L_n$ and $Q_e$ is that the slope of the curve is steep enough for the controller to work in the vicinity of the resonant frequency while changing the gain value between $M_{g\_min}$ and $M_{g\_max}$. From these values the resonant capacitor is set to 68nF and the series inductor is set to 37uH. The magnetizing inductor is the product of $L_n$ and the series inductance and is set to 75uH. For the transformer core we have chosen a planar E core from 3F3 material and an air gap of 200um, which requires 11 turns of primary side windings to achieve 75uH. With ratio of 1.83 this sets the secondary windings to 6. With the total leakage inductance of 6uH, we have winded the series inductor to 28uH.

### III. RESULTS AND DISCUSSION

As designing the resonant converter is a recursive job, a good practice is to simulate the circuit before making it. For our circuit we have chosen a Texas Instruments UCC256404 controller. As TI provides a SIMPLIS model of this controller, we did simulation in the Simplis Test Bench.

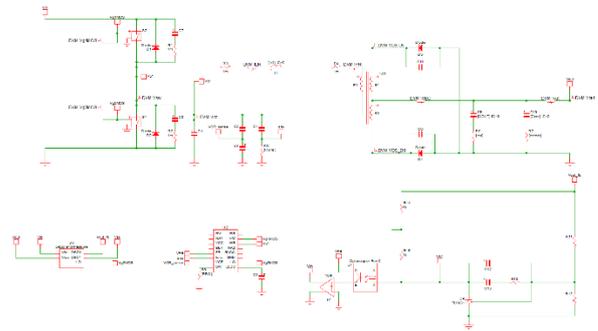

*Figure 6 Simulation model of circuit*

The first simulation is a transient simulation. It calculates a circuit's response over time. The default current is set to 500mA and after 50s there is a 700mA impulse with 10ms duration.

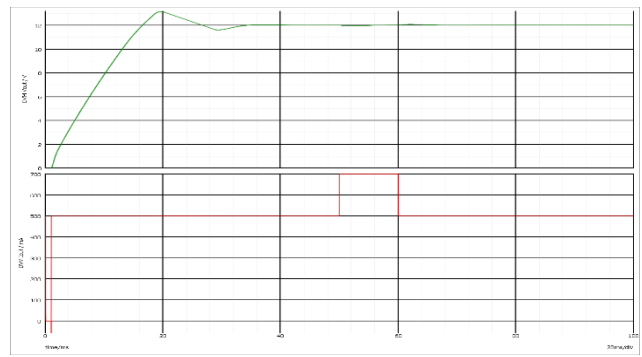

*Figure 7 Simulation transient response*

The second simulation was an AC and a POP - Periodic Operating Point. With an AC simulation we got the frequency response of the controller and the bode plot of the compensation loop. With this we can check stability of the feedback loop.

The POP analysis finds the steady operating point of the circuit and it shows the circuit in the steady state. It also makes other simulations faster because the system already knows the steady state of the circuit.

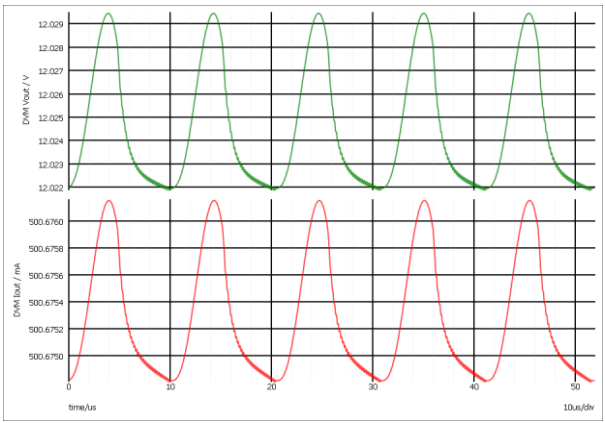

*Figure 8 POP simulation - output voltage (green) and current (red)*

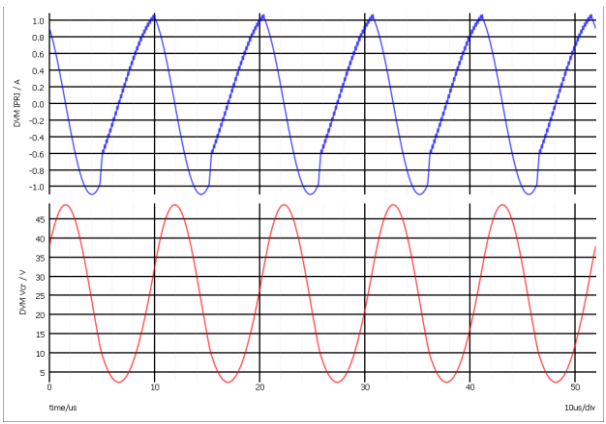

*Figure 9 POP simulation – resonant circuit current (blue), resonant circuit voltage (red)*

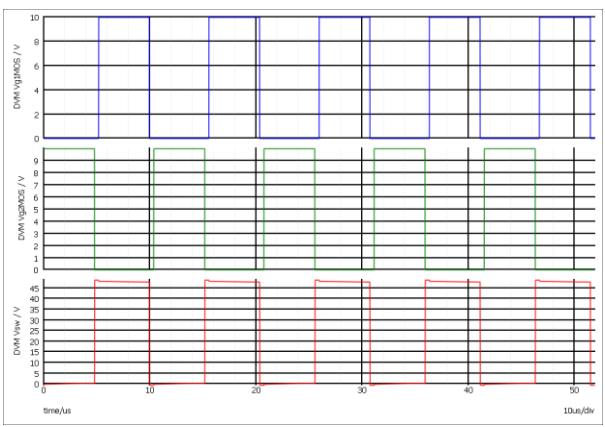

*Figure 10 POP simulation – HS FET gate (blue), LS FET gate (green), switching node (red)*

After simulation, we have produced the board and made the resonant power supply.

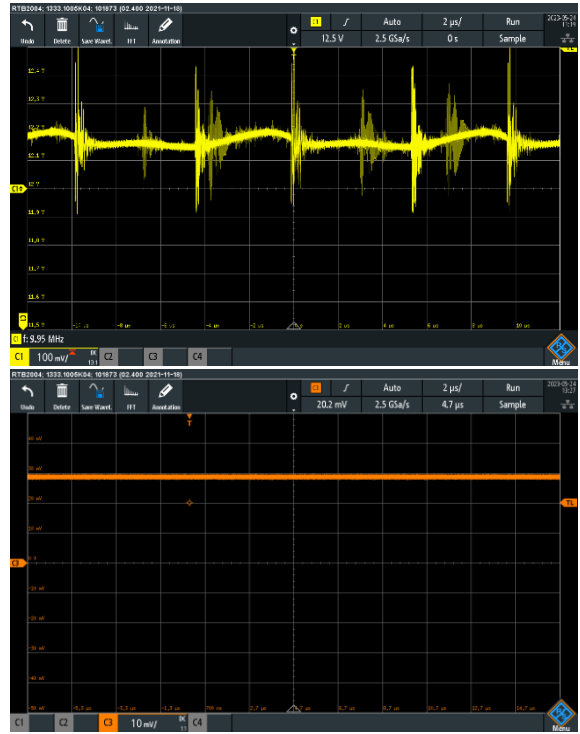

*Figure 11 Measured output voltage and current (10mV/div = 100mA/div)*

Measured output voltage has more ripple than the simulated one. The reason are real components which are not taken into the simulation model and parasitic parameters of the board. This board is just a prototype, in the final version we will add better filtering.

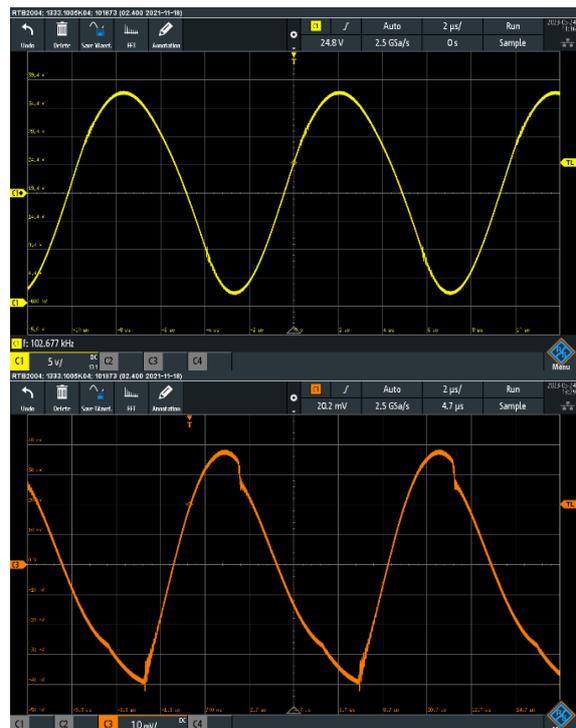

*Figure 12 Measured resonant circuit voltage and current (10mV/div = 200mA/div)*

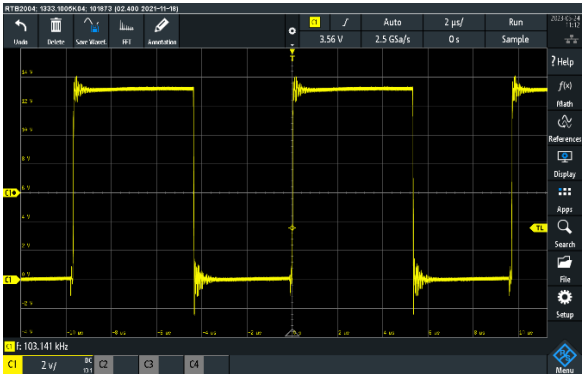

*Figure 13 Measured low side gate signal*

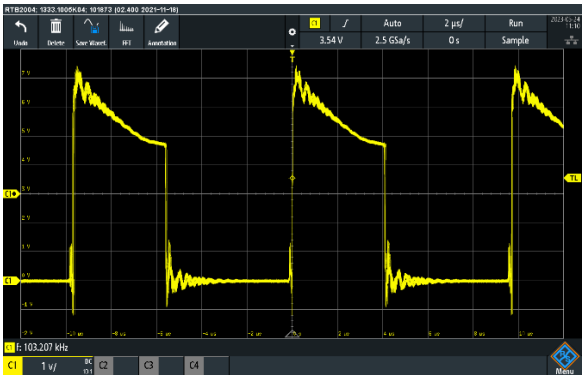

*Figure 14 Measured high side gate signal*

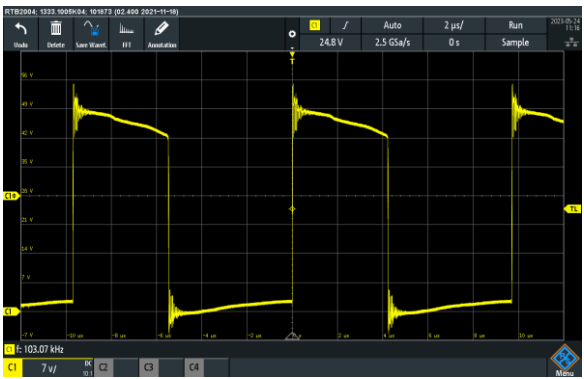

*Figure 15 Measured switching node signal*

Low side gate, high side gate and switching node signals are close to the simulated ones. Measured signals have slightly higher frequency, but it is within the range. The main reason for that is that this measurement is taken on 300mA of the output load, and it requires the lower voltage function gain. When the load gets up to 500mA the switching frequency is near 100kHz. With this, we can see that our load regulation is working well. Another reason for the higher frequency is that our inductor and transformer are not ideal, they are real components and have parasitic components which are not taken into simulation. Although we have measured the inductor and the transformer, they can't be 100% equal to the calculated ones. This means that resonant frequency is not right on 100kHz, but it is in the vicinity of 100kHz. The same reason is for the slightly lower resonant circuit voltage than the simulated one, but the waveform is the same. The current waveform is also the same, only mirrored by the X-axis because the current probe was set differently.

## IV. CONCLUSIONS

LLC converters can be used not only for the high power application, but also for low noise applications. In this paper we designed a low power LLC converter. We have achieved lower noise than the same specifications buck converter. Voltage spikes on the output voltage will be improved in next design, with better layout and filtering. This board is just a prototype and has a lot of dead bug addons as can be seen in Figure 16. The reason for a board like this is because we were using components which were laying around because of long delivery times. This series inductor is over dimensioned for this amount of current, same as the transistors, but we wanted to have the same space if something goes wrong.

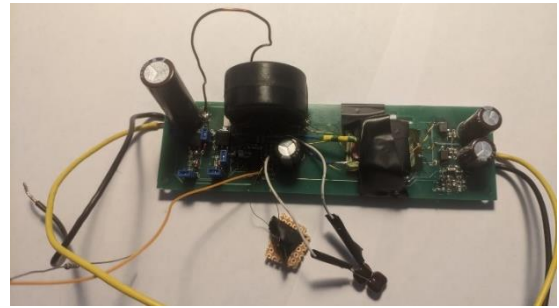

*Figure 16 Produced board*